# Comparison of different forms for the "spin" and "orbital" components of the angular momentum of light*

A. M. Stewart

*Emeritus Faculty, The Australian National University,
Canberra, ACT 0200, Australia.*

*http://grapevine.net.au/~a-stewart/index.html*

### Abstract

We compare three attempts that have been made to decompose the angular momentum of the electromagnetic field into components of an "orbital" and "spin" nature. All three expressions are different, and it appears, on the basis of classical electrodynamics, that there is no preferred way of decomposing the angular momentum of the electromagnetic field into orbital and spin components, even in an inertial frame.

## I. INTRODUCTION

The total angular momentum **J** of the electromagnetic field is given (in SI units) [1] by

$$\mathbf{J}(t) = \varepsilon_0 \int d^3x\, \mathbf{x} \times [\mathbf{E}(\mathbf{x},t) \times \mathbf{B}(\mathbf{x},t)] \qquad (1)$$

Henceforth we will suppress the time coordinate *t* of the fields, all of which depend on time, and also the $\varepsilon_0$ factor.

There has been debate for a long time over whether the total angular moment **J** of the electromagnetic field can be decomposed into an orbital part **L** and a spin part **S** so that

$$\mathbf{J} = \mathbf{L} + \mathbf{S} \qquad (2)$$

Some authors [2] argue that on the basis of first principles it is not possible to do this, others [3,4,5] show that forms can be demonstrated that appear to be, at least algebraically, of a spin and orbital nature.

By means of partial integration Ohanian [4] effected a decomposition with $\mathbf{J} = \mathbf{L}' + \mathbf{S}' + \mathbf{J}'_b$

$$\mathbf{S}' = \int d^3x\, \mathbf{E}(\mathbf{x}) \times \mathbf{A}(\mathbf{x}) \qquad (3)$$

$$\mathbf{L}' = \sum_i \int d^3x\, E^i(\mathbf{x})(\mathbf{x} \times \nabla_x) A^i(\mathbf{x}) \qquad (4)$$

where **A** is the vector potential {$\mathbf{B}(\mathbf{x}) = \nabla_x \times \mathbf{A}(\mathbf{x})$} and $\nabla_x$ is the gradient operator that operates on functions of **x**. Ohanian assumed that the electric charge density $\rho$ was zero and deemed (3) and (4) to be the spin and orbital components respectively of the electromagnetic field on the grounds that the integrand of (3) was not explicitly linear in the **x** coordinate whereas the integrand of (4) was. When the charge density is not zero a bound term $\mathbf{J'}_b$, considered also to be of an orbital nature

$$\mathbf{J'}_b = \int d^3 x \, \mathbf{x} \times \mathbf{A}(\mathbf{x}) \nabla_x \cdot \mathbf{E}(\mathbf{x}) = \int d^3 x \, \mathbf{x} \times \mathbf{A}(\mathbf{x}) \rho(\mathbf{x}) \tag{5}$$

is obtained on whose form all writers agree [6]. Although the sum of (3) and (4) and (5) $\mathbf{J} = \mathbf{L'} + \mathbf{S'} + \mathbf{J'}_b$ is gauge invariant, the individual terms are not and so have no physical interpretation until the gauge of the vector potential is fixed completely.

In the absence of electric charge density, so that the sum of (3) and (4) is gauge invariant, a gauge transformation of the form $\mathbf{A} \to \mathbf{A} + \nabla \chi$ leads to a change of (3) by $\Delta \mathbf{S'} = \int d^3 x \, \mathbf{E}(\mathbf{x}) \times \nabla \chi$ and a change of (4) by $\Delta \mathbf{L'} = \sum_i \int d^3 x \, E^i(\mathbf{x})(\mathbf{x} \times \nabla) \partial \chi / \partial x^i$. A partial integration shows that these terms are of equal magnitude and opposite sign so that they sum to zero. Therefore, by suitably choosing the gauge function, a gauge transformation has the effect of shifting weight from (3) to (4) or *vice versa*. A variety of expressions for (3) and (4) can be obtained by making a specific gauge transformation and then fixing the gauge. Both of (3) and (4) will then be individually gauge invariant but there is no obvious criterion for deciding which is the best gauge to use and which of all the possible variants of (3) and (4) is best.

Despite this ambiguity, Cohen-Tannoudiji *et al.* [3] chose to use the Coulomb (or transverse) gauge, defined by the gauge condition $\nabla \cdot \mathbf{A}_t = 0$, which gives

$$\mathbf{S''} = \int d^3 x \, \mathbf{E}(\mathbf{x}) \times \mathbf{A}_t(\mathbf{x}) \tag{6}$$

$$\mathbf{L''} = \sum_i \int d^3 x \, E^i(\mathbf{x})(\mathbf{x} \times \nabla_x) A_t^i(\mathbf{x}) \tag{7}$$

$$\mathbf{J''}_b = \int d^3 x \, \rho(\mathbf{x}) \mathbf{x} \times \mathbf{A}_t(\mathbf{x}) \;. \tag{8}$$

Since the Coulomb gauge is a completely fixed gauge and has no remaining gauge arbitrariness (6), (7) and (8) are individually well defined with $\mathbf{J} = \mathbf{L''} + \mathbf{S''} + \mathbf{J''}_b$.

Cohen-Tannoudji *et al.* [3] went further and expanded the electric field **E** into transverse $\mathbf{E}_t = -\dot{\mathbf{A}}_t = -\partial \mathbf{A}_t / \partial t$ and longitudinal $\mathbf{E}_l = -\nabla \phi$ parts, where $\phi$ is the scalar potential with
$\mathbf{E} = \mathbf{E}_t + \mathbf{E}_l$, to get

$$\mathbf{S'''} = \int d^3 x \, \mathbf{E}_t(\mathbf{x}) \times \mathbf{A}_t(\mathbf{x}) = -\int d^3 x \, \dot{\mathbf{A}}_t(\mathbf{x}) \times \mathbf{A}_t(\mathbf{x}) \tag{9}$$

$$\mathbf{L'''} = \sum_i \int d^3 x \, E_t^i(\mathbf{x})(\mathbf{x} \times \nabla_x) A_t^i(\mathbf{x}) = -\sum_i \int d^3 x \, \dot{A}_t^i(\mathbf{x})(\mathbf{x} \times \nabla_x) A_t^i(\mathbf{x}) \;. \tag{10}$$

The bound component remains the same as (8). It will be shown in section **II** of this paper that the terms that involve the scalar potential in (6) and (7) cancel so that $\mathbf{J} = \mathbf{L''} + \mathbf{S''} + \mathbf{J''}_b$ in

(6, 7, 8) and $\mathbf{J} = \mathbf{L}''' + \mathbf{S}''' + \mathbf{J}''_b$ in (9, 10, 8) but $\mathbf{S}''$ in (6) differs from $\mathbf{S}'''$ in (9) and $\mathbf{L}''$ in (7) differs from $\mathbf{L}'''$ in (10). The forms of (9) and (10) have also been used by van Enk and Nienhuis [5]. The general explicit form for $\mathbf{A}_t$, given in Eq. (13), was not specified by these writers. Although the total angular momentum of the field has been shown to satisfy the canonical commutation relations for angular momentum [14] the components have not [5].

On the other hand, Stewart [7] found a decomposition $\mathbf{J} = \mathbf{L}'''' + \mathbf{S}'''' + \mathbf{J}''_b$ from decomposing the electric field with the Helmholtz theorem [8]

$$\mathbf{S}'''' = \frac{1}{4\pi} \int d^3x \int d^3y \frac{\mathbf{B}(\mathbf{x}) \times \dot{\mathbf{B}}(\mathbf{y})}{|\mathbf{x} - \mathbf{y}|} \tag{11}$$

and

$$\mathbf{L}'''' = \frac{1}{4\pi} \int d^3x \int d^3y \, \mathbf{B}(\mathbf{x}) \cdot \dot{\mathbf{B}}(\mathbf{y}) \frac{\mathbf{x} \times \mathbf{y}}{|\mathbf{x} - \mathbf{y}|^3} \, . \tag{12}$$

This decomposition uses the $\mathbf{E}$ and $\mathbf{B}$ fields throughout so no issues of gauge arbitrariness arise. Nor is there any complication that arises from using the longitudinal and transverse components of the fields, as $\mathbf{B}$ is entirely transverse. The expression for $\mathbf{J}''_b$ was again given by (8) [6]. Eqs. (8, 11,12) have been used elsewhere [9] to show that the electromagnetic field makes zero contribution to the angular momentum of the physical electron described by the Lagrangian of quantum electrodynamics and to resolve the paradox concerning the angular momentum of a plane electromagnetic wave [10]. Application has also been made to paraxial rays [11] where it has been shown that the same results are obtained from (11, 12) as from (9, 10).

The question addressed in this paper is how (11) and (12) are related to (9) and (10) when the relation below (13) [12,13] that expresses the vector potential of the Coulomb gauge explicitly in terms of the instantaneous magnetic field is used:

$$\mathbf{A}_t(\mathbf{x}) = \nabla_x \times \int d^3y \frac{\mathbf{B}(\mathbf{y})}{4\pi |\mathbf{x} - \mathbf{y}|} = -\int \frac{d^3y}{4\pi} \mathbf{B}(\mathbf{y}) \times \nabla_x \frac{1}{|\mathbf{x} - \mathbf{y}|} \, . \tag{13}$$

In section **II** of the paper we summarise the derivation [3] leading from (1) to (9) and (10). In section **III** we obtain an expression for $\mathbf{S}'''$ of (9) and $\mathbf{L}'''$ of (10) using (13) and demonstrate that $\mathbf{L}''' + \mathbf{S}''' = \mathbf{L}'''' + \mathbf{S}''''$. Section **IV** summarises the conclusion of the paper that there is, on the basis of classical physics, no unique decomposition of the angular momentum of the electromagnetic field into spin and orbital components, even in a fixed inertial frame.

## II. DERIVATION OF THE STANDARD VERSION OF THE ANGULAR MOMENTUM DECOMPOSITION

To obtain (3) and (4) from (1) we consider the component of the cross product

$$\mathbf{E} \times (\nabla_x \times \mathbf{A})|^i = \varepsilon^{ijk} \varepsilon^{klm} E^j \frac{\partial A^m}{\partial x^l} \, . \tag{14}$$

By multiplying the Levi-Civita symbols we find [4]

$$\mathbf{E} \times (\nabla_x \times \mathbf{A}) = E^n \nabla_x A^n - (\mathbf{E} \cdot \nabla_x) \mathbf{A} \, . \tag{15}$$

Next, consider the identity

$$\frac{\partial}{\partial x^m}(x^j E^m A^k) = \delta_{j,m} E^m A^k + x^j (A^k \frac{\partial E^m}{\partial x^m} + E^m \frac{\partial A^k}{\partial x^m}) \qquad (16)$$

and integrate this over $x^m$ from minus infinity to plus infinity. The left hand side vanishes because the integrand is zero at those points. Integrate over the other two components of **x** to produce a volume integral over all space of $d^3x$ then sum over *m*. The result, when multiplied by $\varepsilon^{ijk}$ and integrated, substituted into the angular momentum obtained from (15) and specialised to the Coulomb gauge, is

$$\mathbf{J} = \int d^3x [\mathbf{x} \times \mathbf{A}_t (\nabla_x \cdot \mathbf{E}) + \mathbf{E} \times \mathbf{A}_t + E^n (\mathbf{x} \times \nabla_x) A_t^n] \qquad (17)$$

This reproduces Eqs. (3, 4, 5) for general gauge and (6, 7, 8) for the Coulomb gauge.

The next step is to decompose the electric field into its longitudinal $-\nabla\phi$ and transverse $\mathbf{E}_t$ components. Since the divergence of a transverse vector field is zero the first term in (17) is unchanged from (8). The term associated with the potential in the third term of (17) is

$$\Delta J^3 |^i = -\sum_{j,k,n} \int d^3x \, \varepsilon^{ijk} \frac{\partial \phi}{\partial x^n} x^j \frac{\partial A_t^n}{\partial x^k} \qquad (18)$$

Consider now the identity

$$\frac{\partial}{\partial x^k}(x^j A_t^n \frac{\partial \phi}{\partial x^n}) = \delta_{j,k} A_t^n \frac{\partial \phi}{\partial x^n} + x^j (\frac{\partial A_t^n}{\partial x^k} \frac{\partial \phi}{\partial x^n} + A_t^n \frac{\partial^2 \phi}{\partial x^k \partial x^n}) \qquad (19)$$

When (19) is put into (18) the term coming from the first term of the right hand side of (19) vanishes because of the product of the Kronecker delta and the Levi Civita functions and we get

$$\Delta J^3 |^i = \sum_{j,k,n} \int d^3x \, \varepsilon^{ijk} x^j A_t^n \frac{\partial^2 \phi}{\partial x^k \partial x^n} \qquad (20)$$

From the identity

$$\frac{\partial}{\partial x^n}(x^j A_t^n \frac{\partial \phi}{\partial x^k}) = \delta_{j,n} A_t^n \frac{\partial \phi}{\partial x^k} + x^j (\frac{\partial A_t^n}{\partial x^n} \frac{\partial \phi}{\partial x^k} + A_t^n \frac{\partial^2 \phi}{\partial x^k \partial x^n}) \qquad , \qquad (21)$$

noting that $\nabla \cdot \mathbf{A}_t = 0$, we find

$$\Delta J^3 |^i = -\sum_{j,k,n} \int d^3x \, \varepsilon^{ijk} \delta_{j,n} A_t^n \frac{\partial \phi}{\partial x^k} = -\sum_{j,k} \int d^3x \, \varepsilon^{ijk} A_t^j \frac{\partial \phi}{\partial x^k} \qquad . \qquad (22)$$

The angular momentum of the second term of (17) arising from the potential is

$$\Delta J^2 \mid^i = -\sum_{j,k} \int d^3x \, \varepsilon^{ijk} \frac{\partial \phi}{\partial x^j} A_t^k \tag{23}$$

Noting the order of the indices, Eqs. (22) and (23) sum to zero so we find that the contribution of the longitudinal electric field to the total angular momentum is zero so **J** = **L**''' + **S**''' + **J**''$_b$ with the components **L**''' and **S**''' given by Eqs. (9) and (10). We see already that the angular momentum of the electromagnetic field can be expressed in two different ways, as **J** = **L**''' + **S**''' + **J**''$_b$, given by (6), (7) and (8) and as **J** = **L**'' + **S**'' + **J**''$_b$ given by equations (9), (10) and (8).

### III. CONFIRMATION OF THE VALIDITY OF THE NEW VERSION OF THE ANGULAR MOMENTUM DECOMPOSITION

In this section of the paper we substitute the expression for the vector potential of the Coulomb gauge (13) into (9) and (10) in order to find how (9) and (10) are related to (11) and (12). We will find that the sum of (9) and (10) is equal to the sum of (11) and (12). First we examine **L**''' of (10). This may be expressed as

$$\begin{aligned} \mathbf{L}''' &= -\int d^3x \, \mathbf{x} \times \nabla_x [\dot{\mathbf{A}}_t(\mathbf{z}) \cdot \mathbf{A}_t(\mathbf{x})]\big|_{z=x} \\ &= -\int d^3x \, \mathbf{x} \times \{\dot{\mathbf{A}}_t(\mathbf{x}) \times \mathbf{B}(\mathbf{x}) + [\dot{\mathbf{A}}_t(\mathbf{x}) \cdot \nabla_\mathbf{x}] \mathbf{A}_t(\mathbf{x})\} \end{aligned} \tag{24}$$

Using (13), the first term of (24) comes to

$$\mathbf{L}'''^1 = -\frac{1}{4\pi} \int d^3x \int d^3y \, \mathbf{x} \times \{\mathbf{B}(\mathbf{x}) \times [\dot{\mathbf{B}}(\mathbf{y}) \times \nabla_x \frac{1}{|\mathbf{x}-\mathbf{y}|}]\} \tag{25}$$

and, by multiplying out the triple vector product, this becomes

$$\mathbf{L}'''^1 = -\frac{1}{4\pi} \int d^3x \int d^3y \, \mathbf{x} \times \{\dot{\mathbf{B}}(\mathbf{y})[\mathbf{B}(\mathbf{x}) \cdot \nabla_x \frac{1}{|\mathbf{x}-\mathbf{y}|}] - \nabla_x \frac{1}{|\mathbf{x}-\mathbf{y}|}[\dot{\mathbf{B}}(\mathbf{y}) \cdot \mathbf{B}(\mathbf{x})]\}. \tag{26}$$

When the gradient in the second term of (26), **L**'''$^{12}$, is expressed explicitly, the term becomes

$$\mathbf{L}'''^{12} = \frac{1}{4\pi} \int d^3x \int d^3y \, \dot{\mathbf{B}}(\mathbf{y}) \cdot \mathbf{B}(\mathbf{x}) \frac{\mathbf{x} \times \mathbf{y}}{|\mathbf{x}-\mathbf{y}|^3} \tag{27}$$

The first term in (26) is expressed in components as

$$\mathbf{L}'''^{11} \mid^i = -\frac{1}{4\pi} \sum_{j,k,m} \varepsilon^{ijk} \int d^3x \int d^3y \, x^j \dot{B}^k(\mathbf{y}) B^m(\mathbf{x}) \frac{\partial}{\partial x^m} \frac{1}{|\mathbf{x}-\mathbf{y}|} \tag{28}$$

By considering the identity

$$\frac{\partial}{\partial x^m}\left[\frac{\dot{B}^k(\mathbf{y}) x^j B^m(\mathbf{x})}{|\mathbf{x}-\mathbf{y}|}\right] = \dot{B}^k(\mathbf{y})\{\delta_{j,m}\frac{B^m(\mathbf{x})}{|\mathbf{x}-\mathbf{y}|} + x^j[B^m(\mathbf{x})\frac{\partial}{\partial x^m}\frac{1}{|\mathbf{x}-\mathbf{y}|} + \frac{1}{|\mathbf{x}-\mathbf{y}|}\frac{\partial B^m(\mathbf{x})}{\partial x^m}]\} \tag{29}$$

we obtain, with $\nabla \cdot \mathbf{B} = 0$,

$$\mathbf{L}'''^{11}|^i = \frac{1}{4\pi} \sum_{j,k,m} \varepsilon^{ijk} \int d^3x \int d^3y \, \dot{B}^k(\mathbf{y}) \frac{\delta_{jm} B^m(\mathbf{x})}{|\mathbf{x}-\mathbf{y}|} \tag{30}$$

or

$$\mathbf{L}'''^{11} = \frac{1}{4\pi} \int d^3x \int d^3y \, \frac{\mathbf{B}(\mathbf{x}) \times \dot{\mathbf{B}}(\mathbf{y})}{|\mathbf{x}-\mathbf{y}|} \tag{31}$$

The second term $\mathbf{L}'''^2$ of (24) is simplified by writing it in components and using the relation

$$\frac{\partial}{\partial x^m}(x^j \dot{A}_t^m A_t^k) = \delta_{jm} \dot{A}_t^m A_t^k + x^j \left( \frac{\partial \dot{A}_t^m}{\partial x^m} A_t^k + \dot{A}_t^m \frac{\partial A_t^k}{\partial x^m} \right) \tag{32}$$

with the Coulomb gauge condition $\nabla \cdot \mathbf{A}_t = 0$ to get

$$\mathbf{L}'''^2 = \int d^3x \, \dot{\mathbf{A}}_t(\mathbf{x}) \times \mathbf{A}_t(\mathbf{x}) \tag{33}$$

Accordingly, $\mathbf{L}''' = \mathbf{L}'''^{11} + \mathbf{L}'''^{12} + \mathbf{L}'''^2$ and

$$\begin{aligned}\mathbf{L}''' &= \frac{1}{4\pi} \int d^3x \int d^3y \, \frac{\mathbf{B}(\mathbf{x}) \times \dot{\mathbf{B}}(\mathbf{y})}{|\mathbf{x}-\mathbf{y}|} + \frac{1}{4\pi} \int d^3x \int d^3y \, \dot{\mathbf{B}}(\mathbf{y}) \cdot \mathbf{B}(\mathbf{x}) \frac{\mathbf{x} \times \mathbf{y}}{|\mathbf{x}-\mathbf{y}|^3} \\ &+ \int d^3x \, \dot{\mathbf{A}}_t(\mathbf{x}) \times \mathbf{A}_t(\mathbf{x})\end{aligned} \tag{34}$$

Since $\mathbf{S}'''$ of (9) is given by the negative of the last term of (34) it follows that $\mathbf{L}''' = \mathbf{S}'''' + \mathbf{L}'''' - \mathbf{S}'''$ and so $\mathbf{L}''' + \mathbf{S}''' = \mathbf{L}'''' + \mathbf{S}''''$. We find that both schemes give the same total angular momentum.

## IV. CONCLUSION

In this paper we have compared three different expressions for the angular momentum of the electromagnetic field; they are (35, 36, 37):

$$\mathbf{J} = \int d^3x \, [\mathbf{x} \times \mathbf{A}_t \, \rho(\mathbf{x}) + \varepsilon_0 \mathbf{E} \times \mathbf{A}_t + \varepsilon_0 E^n (\mathbf{x} \times \nabla_x) A_t^n] \tag{35}$$

$$\mathbf{J} = \int d^3x \, [\mathbf{x} \times \mathbf{A}_t \, \rho(\mathbf{x}) - \varepsilon_0 \dot{\mathbf{A}}_t \times \mathbf{A}_t - \varepsilon_0 \dot{A}_t^n (\mathbf{x} \times \nabla_x) A_t^n] \tag{36}$$

$$\begin{aligned}\mathbf{J} &= \int d^3x \, \mathbf{x} \times \mathbf{A}_t(\mathbf{x}) \rho(\mathbf{x}) + \frac{\varepsilon_0}{4\pi} \int d^3x \int d^3y \, \frac{\mathbf{B}(\mathbf{x}) \times \dot{\mathbf{B}}(\mathbf{y})}{|\mathbf{x}-\mathbf{y}|} \\ &+ \frac{\varepsilon_0}{4\pi} \int d^3x \int d^3y \, \mathbf{B}(\mathbf{x}) \cdot \dot{\mathbf{B}}(\mathbf{y}) \frac{\mathbf{x} \times \mathbf{y}}{|\mathbf{x}-\mathbf{y}|^3}\end{aligned} \tag{37}$$

where the vector potential of the Coulomb gauge $\mathbf{A}_t$ is given by (13) and the $\varepsilon_0$ factor has been restored. The expressions all have the algebraic form of the sum of a spin and an orbital component and each component of all three expressions is gauge invariant. However equation (37) is gauge covariant, it has the same form in every gauge, whereas (35) and (36) are not; they are made gauge invariant by the artificial procedure of fixing the gauge to be the Coulomb gauge. All three expressions are different in form and, on the basis of classical electrodynamics, there seems to be no reason to prefer one to another. The last form (37) [7] has the pedagogical advantage of being derived in a manifestly gauge-invariant manner entirely in terms of fields through the Helmholtz theorem; the gauge ambiguity referred to in section **I** does not arise in its derivation. It appears that, in classical electromagnetism, there is no preferred way of decomposing the angular momentum of the field into orbital and spin components, even in an inertial frame.

**References**

*An earlier version of this paper has been published in the International Journal of Optics, Volume 2011, article ID 728350 (2011). doi:10.1155/2011/7828350 http://www.hindawi.com/journals/ijo/2011/728350/.

[1] J. D. Jackson, *Classical Electrodynamics*, 3rd ed. (Wiley, New York, 1999).
[2] W. Heitler, *The Quantum Theory of Radiation*, 3rd ed. (Clarendon Press, Oxford, 1954).
[3] C. Cohen-Tannoudji, J. Dupont-Roc, and G. Gilbert, *Photons and Atoms*, (Wiley, New York, 1989).
[4] H. C. Ohanian, *What is spin?*, American Journal of Physics **54**(6), 500-505 (1986).
[5] S. J. van Enk and G. Nienhuis, *Commutation rules and eigenvalues of spin and orbital angular momentum of radiation fields*, Journal of Modern Optics **41**(5), 963-977 (1994).
[6] A. M. Stewart, *Equivalence of two mathematical forms for the bound angular momentum of the electromagnetic field*, Journal of Modern Optics **52**(18), 2695-2698 (2005). arXiv:physics/0602157v3 [physics.optics]
[7] A. M. Stewart, *Angular momentum of light*, Journal of Modern Optics **52**(8), 1145-1154 (2005). arXiv:physics/0504078v2 [physics.optics]
[8] A. M. Stewart, *Longitudinal and transverse components of a vector field*, (2009). arXiv:0801.0335v2 [physics.class-ph]
[9] A. M. Stewart, *Angular momentum of the physical electron*, Canadian Journal of Physics **87**(2), 151-152 (2009). arXiv:quant-ph/0701136v2
[10] A. M. Stewart, *Angular momentum of the electromagnetic field: the plane wave paradox resolved*, European Journal of Physics **26**, 635-641 (2005). arXiv:physics/0504082v3 [physics.class-ph]
[11] A. M. Stewart, *Derivation of the paraxial form of the angular momentum of the electromagnetic field from the general form*, Journal of Modern Optics **53**(13), 1947-1952 (2006). arXiv:physics/0608029v2 [physics.class-ph]
[12] A. M. Stewart, *Vector potential of the Coulomb gauge*, European Journal of Physics **24**, 519-524 (2003).
[13] A. M. Stewart, *Reply to Comments on 'Vector potential of the Coulomb gauge'*, European Journal of Physics **25**, L29-L30 (2004). arXiv:1005.5502v1 [physics.class-ph]
[14] A. M. Stewart. *Quantization of the free electromagnetic field implies quantization of its angular momentum,* Journal of Modern Optics **58**(9), 772-776 (2011). doi:10.1080/09500340.2011.571797 http://arxiv.org/abs/1011.4131